\documentclass[11pt, a4paper]{article}
\pdfoutput=1
\usepackage{jcappub}
\usepackage[utf8]{inputenc}
\usepackage[T1]{fontenc}
\newcommand{\la}{\lambda_1}
\newcommand{\lb}{\lambda_2}
\newcommand{\lc}{\lambda_3}
\newcommand{\g}{\,\mathrm{GeV}}
\newcommand{\be}{\begin{equation}}
\newcommand{\ee}{\end{equation}}

\newcommand{\Vtree}{V^{(0)}}
\newcommand{\Vone}{V^{(1)}}
\newcommand{\VT}{V^{T}}
\newcommand{\f}{\varphi}
\newcommand{\ms}{\overline{\textrm{MS}}}

\newcommand{\mgw}{\mu_{\textrm{GW}}}

\usepackage{amsthm}
\usepackage{amsfonts}
\usepackage{physics}
\usepackage[mathscr]{eucal}
\allowdisplaybreaks[3]
\usepackage{slashed}
\title{Gravitational waves from conformal symmetry breaking}
\author{Tomislav Prokopec,}
\author{Jonas Rezacek,}
\author{Bogumi\l a~\'Swie\.zewska}
\affiliation{Institute for Theoretical Physics, Spinoza Institute \& EMME$\Phi$, Utrecht University,\\Princetonplein 5, 3584 CC Utrecht, The Netherlands}

\emailAdd{t.prokopec@uu.nl,  j.d.rezacek@students.uu.nl, b.swiezewska@uu.nl}

\abstract{
We consider the electroweak phase transition in the conformal extension of the standard model known as SU(2)cSM. Apart from the 
standard model particles, this model contains an additional scalar and a gauge field that are both charged under the hidden SU(2)$_X$. 
This model generically exhibits a very strong phase transition that proceeds after a large amount of supercooling. We estimate the gravitational wave spectrum produced  in this model and show that its amplitude
and frequency fall within the observational window of LISA. We also discuss potential pitfalls and relevant points of improvement
required to attain reliable estimates of the gravitational wave production in this --- as well as in more general --- class of models.
In order to improve perturbativity during the early stages of transition that ends with bubble nucleation, we solve 
a thermal gap equation in the scalar sector inspired by the 2PI effective action formalism.\note{This is the Accepted Manuscript version of an article accepted for publication in Journal of Cosmology and Astroparticle Physics. Neither SISSA Medialab Srl nor IOP Publishing Ltd is responsible for any errors or omissions in this version of the manuscript or any version derived from it. The Version of Record is available online at \url{https://doi.org/10.1088/1475-7516/2019/02/009}.}

}

\arxivnumber{1809.11129}

\begin{document}
\maketitle

\section{Introduction}

Most of the elementary particles that we know are massive. This implies that the gauge symmetry of the Standard Model (SM) must be broken, since the symmetry forbids mass terms for gauge bosons. However, taking into account the thermal evolution of the Universe from the hot Big Bang to the cold present times, we are brought to the conclusion that the Universe must have started in a symmetric state where the symmetry was preserved and all particles were massless~\cite{Kirzhnits:1972, Weinberg:1974, Dolan:1974}. Subsequently, it underwent a phase transition during which a nonvanishing vacuum expectation value (VEV) for a scalar field was developed and masses for other particles were generated through the mechanism of spontaneous symmetry breaking. This is typically realised via the Brout--Englert--Higgs mechanism, which assumes a negative mass term for the scalar field that is not excluded by the gauge symmetry and induces a nontrivial minimum of the potential. In the present paper we entertain the idea that the Universe might have started in a state with an enhanced symmetry --- not only the gauge symmetry was present but also a (classical) conformal symmetry was realised, prohibiting any massive parameters in the lagrangian.  Then the conformal symmetry must have been broken by quantum effects, generating a nontrivial VEV of the scalar field(s) and mass terms for all the massive particles.\footnote{In the present work we do not address neutrino masses, for some approaches see e.g.~\cite{Khoze:2013-2, Karam:2015, Plascencia:2016, Brdar:2018}.}

Two aspects of such a phase transition in the early Universe are of physical importance. First, if the phase transition is of first order and sufficiently strong it can constitute a necessary ingredient for a successful baryogenesis. Second, such a violent phenomenon can leave an imprint in the spacetime structure in the form of gravitational waves (GW)  that can prevail to the present times and possibly be observed. The GW signal from the conformal and electroweak symmetry breaking phase transition is the main interest of the present article. 

Lots of work has been devoted to modelling the GW spectra, see refs.~\cite{Turner:1990, Hogan:1986, Kosowsky:1992, Kosowsky:1991, Kosowsky:1992-2, Kamionkowski:1993} for the early results on gravitational waves from phase transitions, refs.~\cite{Apreda:2001, Nicolis:2003, Grojean:2006, Caprini:2007, Huber:2008, Hindmarsh:2013, Hindmarsh:2015, Hindmarsh:2017} for subsequent developments, ref.~\cite{Caprini:2015} for a recent review. These results have been used to analyse GW in various models featuring strong first order phase transition, see ref.~\cite{Kanemura:2018} for a rich list of references. GWs from phase transition have been studied within models where the conformal symmetry is broken due to Coleman-Weinberg mechanism (perturbative dynamics)~\cite{Espinosa:2008, Dorsch:2014, Jaeckel:2016, Hashino:2016, Jinno:2016, Marzola:2017} and also in the context of models with approximately conformal potentials with strong dynamics or extra dimensions, see e.g.~\cite{Randall:2006, Konstandin:2010, Konstandin:2011, Kubo:2016, vonHarling:2017, Bruggisser:2018}.

The idea of underlying conformal symmetry is attractive since it allows to abandon the arbitrary negative mass parameter for the scalar field. It is produced dynamically and the same time alleviating the hierarchy problem~\cite{Bardeen:1995}. Since the very minimal conformal SM (cSM) is not capable of explaining the observed particles' masses~\cite{Coleman:1973},\footnote{In ref.~\cite{Coleman:1973} the top quark was not taken into account since it had not yet been discovered at that time, which led the authors to the conclusion that in cSM a stable radiatively generated minimum can be formed, however the predicted mass of the Higgs boson was too low. After inclusion of the top quark, the conclusion is changed and it can be shown that there is no stable nontrivial minimum in cSM.} conformal models need some field content that goes beyond the SM. In this way they can also attempt to model phenomena like dark matter, baryogenesis, neutrino masses (see e.g.\ refs.~\cite{Meissner:2006, Meissner:2008, Foot:2007s, AlexanderNunneley:2010, Farzinnia:2013, Gabrielli:2013, Allison:2014, Sannino:2015, Ghorbani:2017,Foot:2007-3,Espinosa:2007, Espinosa:2008, Lee:2012, Helmboldt:2016, Fink:2018, Davoudiasl:2014,Hempfling:1996,Foot:2007,Chang:2007, Hambye:2008, Iso:2009, Dermisek:2013, Hambye:2013, Carone:2013, Khoze:2013-1, Khoze:2013-2,Khoze:2013-3,Heikinheimo:2013,Khoze:2014, Altmannshofer:2014, Benic:2014,  Karam:2015, Karam:2016, Plascencia:2015, DiChiara:2015, Plascencia:2016, Oda:2017, Guo:2015, Hashino:2015, Hashino:2016, Loebbert:2018, Hambye:2018, Chataignier:2018RSB}). In this paper we focus on the SU(2)cSM model. It consists of the cSM extended by a new sector composed of a scalar field and a new gauged symmetry group SU(2)$_X$. The two sectors are only coupled through a scalar portal coupling, which means that the new scalar is a singlet of the SM symmetry group, while being a doublet of the new SU(2)$_X$. This model has been analysed in refs.~\cite{Hambye:2013, Carone:2013, Khoze:2014, Hambye:2018}  (also in an extended form in refs.~\cite{Karam:2015, Plascencia:2016}) and more recently the radiative symmetry breaking (RSB) in this model has been thoroughly analysed in ref.~\cite{Chataignier:2018RSB}. It was shown that via RSB particle spectrum consistent with observations can be modelled within SU(2)cSM at zero temperature. Moreover, the model considered here has few free parameters, while offering interesting phenomenology including a viable candidate for dark matter~\cite{Hambye:2013, Carone:2013, DiChiara:2015,Hambye:2018}. Furthermore, the model is perturbative and stable up to the Planck scale.

In the present paper we study temperature-dependent effective potential and focus on the aforementioned symmetry-breaking phase transition in the early Universe. We find that the signal predicted by the SU(2)cSM is very strong, well within the sensitivity range of the future GW detector LISA~\cite{LISA}.\footnote{Preliminary results on which this paper is based were presented in the master thesis of one of the authors, see ref.~\cite{Rezacek:2018}.}  Thus,  it is an extremely interesting case, also to study the interplay between the gravitational-wave and the particle phenomenology. We therefore  advocate the necessity of providing predictions of this model as reliably as possible to be confronted with the experimental data once it is available.

In the present paper we make a step towards improving the reliability of the theoretical approach to phase transitions in SU(2)cSM.  First of all,  we study the dynamics of the phase transition in two-dimensional field space, allowing the transition to proceed along any direction. This allows us to find full two-dimensional bubble profiles and verify the applicability of the commonly used approaches --- the Gildener-Weinberg approach~\cite{Gildener:1976} which only focuses on the direction between the origin in the field space and the location of the minimum, and the sequential approach, which considers symmetry breaking in the direction of the new scalar filed prior to the symmetry breaking in the SM Higgs sector. Second, it is well known that in the vicinity of a phase transition perturbativity of the loop expansion of the effective potential is jeopardised by large corrections coming mostly from the longitudinal gauge bosons and scalars. Perturbativity can be restored through resummation of the problematic terms. This is usually implemented by inclusion of the thermal masses obtained from resummation of the daisy diagrams. With the aim of improving the accuracy of obtained results as well as widening the region of applicability of our formalism, we make a first step towards resummations  based on the gap equation inspired by the 2PI approach to the effective action~\cite{Cornwall:1974, Berges:2004, AmelinoCamelia:1992, AmelinoCamelia:1993, Funakubo:2012, Laine:2017}, which includes also subleading terms and is valid in the full temperature range.

We start from introducing the SU(2)cSM and defining the effective potential that we use in section~\ref{sec:eff-potential}. Next, in section~\ref{sec:symmetry-breaking} we review the physical phenomena related to the symmetry-breaking transition. Section~\ref{sec:results} is devoted to the analysis of the details of the phase transition and the resulting GW spectra. Motivated by the properties of the phase transition found in section~\ref{sec:results}, in section~\ref{sec:improved-potential} we introduce the gap-equation improvement of the effective potential and analyse the impact thereof on the obtained results. Finally, we conclude in section~\ref{sec:conclusions}.

\section{SU(2)cSM and the effective potential\label{sec:eff-potential}}

The basic tool for studying symmetry breaking is the effective potential, global minimum of which constitutes the vacuum state of the theory. The knowledge of the vacuum state, as well as of the local minima of the effective potential and their temperature dependence is necessary for the analysis of the phase transitions. In this section we introduce the zero- and finite-temperature effective potential. We start from defining the model under scrutiny, the SU(2)cSM.

\subsection{SU(2)cSM}

The basic cSM has been proven incapable of generating a stable minimum~\cite{Coleman:1973}, while the minimal extension of the cSM by a scalar singlet is not capable of reproducing the correct Higgs mass, without introducing large scalar couplings which lead to appearance of Landau poles close to the electroweak scale. Therefore, in this work we study a classically conformal model with extended scalar and gauge sectors, the so-called SU(2)cSM~\cite{Hambye:2013, Carone:2013, Khoze:2014}. The cSM is supplemented with an additional scalar field, which is a singlet under the SM gauge group and a doublet of a new gauge group, SU(2)$_X$. This new group acts trivially on the SM sector.
 This model is minimal in the sense that it is the simplest model that extends the standard model such to retain perturbativity
up the the Planck scale without a significant amount of fine tuning of parameters~\cite{Chataignier:2018,Chataignier:2018RSB}.\footnote{In models with the U(1) gauge group, generically the coupling constant develops a Landau pole. Nonetheless, for certain choices of parameters models based on U(1) symmetry can also be stable up to the Planck scale, see e.g.\ ref.~\cite{Khoze:2014,Loebbert:2018} or refs.~\cite{Hashino:2018, Duch:2015} where similar non-conformal models were considered.}
Of course, one could study more baroque models, in which e.g.\ the hidden group SU(2)$_X$ is generalised to SU(N)$_X$~\cite{DiChiara:2015}
or hidden sector fermions are added, see e.g.~\cite{Croon:2018}.

In SU(2)cSM the interactions between scalars are dictated by the scalar potential
\be
\Vtree(\Phi, \Psi)=\la \left(\Phi^{\dagger}\Phi\right)^2 + \lb \left(\Phi^{\dagger}\Phi\right) \left(\Psi^{\dagger}\Psi \right)+ \lc \left(\Psi^{\dagger}\Psi\right)^2,\notag
\ee
where $\Phi$ is the SM scalar doublet, while $\Psi$ is a doublet under the hidden SU(2)$_X$ gauge group. The boundedness of the potential from below, which is necessary for a stable vacuum state to exist, imposes the following tree-level constraints on the couplings~\cite{Kannike:2012} 
\be
\la \geqslant 0, \quad \lc\geqslant 0,\quad \lb\geqslant -2\sqrt{\la\lc}.\nonumber
\ee
As explained in refs.~\cite{Sher:1989, Chataignier:2018}, if the potential beyond tree level is considered, these constrains should be imposed on the running couplings evaluated at some high energy scale (e.g.\ the Planck scale).

Radiative symmetry breaking in SU(2)cSM has been analysed recently in ref.~\cite{Chataignier:2018RSB} and also discussed earlier in refs.~\cite{Hambye:2013, Carone:2013, Khoze:2014} (also in extended version in refs.~\cite{Karam:2015, Plascencia:2016}). It has been shown that in the SU(2)cSM, in a wide range of the parameter space, RSB can produce correct particle spectrum, i.e.\ the masses of the Higgs boson and the other SM particles are reproduced with perturbative values of the coupling constants. In the present analysis we focus on the parameter space discussed in ref.~\cite{Chataignier:2018RSB}, namely $\lambda_2\in(-0.01,-0.001)$ and $g_X\in(0.1, 1.1)$ ($g_X$ is the coupling constant of the SU(2)$_X$ gauge group) and assume that the Higgs boson is lighter than the other scalar particle. In this case, the VEV of the new scalar field is typically $\mathcal{O}(10)$ times greater than the SM VEV, $v=246\g$. The new gauge boson, denoted by $X$, acquires a mass up to around 2~TeV.\footnote{Smaller absolute values of $\lb$ can be considered, as was done in e.g.~\cite{Carone:2013, Baldes:2018}, in this case the VEV of the new scalar and the mass of the gauge boson $X$ are increased.} It can play a role of a dark matter candidate~\cite{Hambye:2013, Carone:2013, DiChiara:2015,Hambye:2018}.
 The Higgs boson mixes with a new scalar particle and the mixing angle is within experimental bounds in a significant portion of the parameter space.  This, combined with the fact that the new scalar and vector particles are heavier than the Higgs boson, thus do not induce new Higgs decay channels, shows that the Higgs boson should have SM-like phenomenology.

\subsection{Effective potential at zero temperature\label{sec:eff-pot-0-T}}

The effective potential at zero temperature is used to determine the physical predictions that can be observed at present times. It is a classical function of the so-called background (or classical, or average) fields which correspond to vacuum expectation values (VEVs) of the quantum  fields' operators in the presence of an external  source and the class of considered fields is limited to those constant in space and time. The effective potential shares the symmetry of the classical one~\cite{Peskin} therefore by means of the weak SU(2) and the hidden SU(2)$_X$ symmetry we can reduce the effective potential to a function of two scalar fields, which correspond to the background fields of the radial components of the scalar doublets (with a slight abuse of the notation we do not differentiate between the quantum and the background fields since from now on we will only deal with the background fields)
\be
h^2=2\Phi^{\dagger}\Phi, \quad \f^2=2 \Psi^{\dagger}\Psi.\notag
\ee
With these definitions the zeroth order effective potential reads
\be
\Vtree(h,\f)=\frac{1}{4}\left(\la h^4 + \lb h^2 \f^2 + \lc \f^4\right).\label{eq:Vtree}
\ee
The one-loop zero-temperature effective potential is given by the standard formula~\cite{Coleman:1973} (in the $\ms$ scheme and using Landau gauge)
\be
\Vone(h,\f)=\frac{1}{64 \pi^2}\sum_{a}n_a m_a^4(h,\f)\left[\log\frac{m_a^2(h,\f)}{\mu^2}-C_a\right], \label{eq:one-loop}
\ee
where $n_a$ counts the number of degrees of freedom and for a particle of spin $s_a$ is given by
\be
n_a=(-1)^{2s_a} Q_a N_a (2s_a+1),\notag
\ee
where $N_a$ stands for the number of colours and $Q_a=1,2$ for neutral/charged particles. $m_a(h,\f)$ correspond to (tree-level) field-dependent masses. In the scalar sector they correspond to the eigenvalues of the Hessian of the tree-level potential
\begin{align}
\begin{pmatrix}
m_{hh}^2 & m_{h \varphi}^2 \\ m_{\varphi h}^2 & m_{\varphi \varphi}^2
\end{pmatrix} =
\begin{pmatrix}
  3 \lambda_1 h^2 + \frac{\lambda_2}{2} \varphi^2 && \lambda_2 h \varphi\\
 \lambda_2 h \varphi && 3 \lambda_3 \varphi^2 + \frac{\lambda_2}{2} h^2 
\end{pmatrix}
\label{eq:tree-level-mass-matrix}
\end{align}
 and are given by the following formulas
\begin{align}
m^2_{\pm}& = \frac{1}{2}\left(3 \lambda_1 + \frac{\lambda_2}{2}\right)h^2 + \frac{1}{2}\left(\frac{\lambda_2}{2} + 3\lambda_3\right)\varphi^2 \notag\\[2pt]
& \ \ \ \pm\frac{1}{2}\sqrt{\left[\left(3\lambda_1-\frac{\lambda_2}{2}\right)h^2-\left(3\lambda_3-\frac{\lambda_2}{2}\right)\varphi^2\right]^2+4\lambda_2^2h^2\varphi^2}, \label{eq:scalar-masses-tree}\\
m^2_{G,G^{\pm}}&= \lambda_1 h^2 + \frac{\lambda_2}{2}\varphi^2, \\
m^2_{\tilde{G},\tilde{G}^{\pm}} &= \frac{\lambda_2}{2}h^2+\lambda_3 \varphi^2.
\end{align}
For the gauge bosons and the top quark the tree-level field-dependent masses are given by
\be
m_W(h)=\frac{gh}{2},\quad m_Z(h)=\frac{\sqrt{g^2+g'^2}h}{2},\quad m_X(\f)=\frac{g_{X}\f}{2}, \quad m_t(h)=\frac{y_t h}{\sqrt{2}},\notag
\ee
where $g$, $g'$ and $g_X$ are the SU(2), U(1) and the new SU(2)$_X$ gauge couplings, and $y_t$ denotes the top Yukawa coupling. In practical computations we ignore the Goldstone-bosons' contributions to the one-loop correction as their numerical impact on the results is negligible for the values of the couplings considered here (see~\cite{Chataignier:2018RSB}), while including them one has to face several complications. The Goldstone masses squared at certain field values become negative, thus introducing imaginary component to the effective potential. Moreover,  due to Goldstone contributions infrared divergences appear~\cite{Martin:2014, Elias-Miro:2014}. Last but not least, if one computes the effective potential in an arbitrary Fermi gauge (see e.g.~\cite{Andreassen:2014, Espinosa:2016, Loebbert:2018}), it turns out that the gauge dependence is exclusively related to the Goldstone (and ghost) contributions, thus neglecting the Goldstone contributions leaves the effective potential independent of the $\xi$ parameter.
\subsection{Effective potential at finite temperature\label{sec:eff-pot-finite-T}}
In order to learn how the properties of the Universe evolved as it cooled down after the Big Bang, one needs to extend the effective potential (and field theory in general) to contain the notion of temperature~\cite{Weinberg:1974, Dolan:1974} (see also ref.~\cite{Quiros:1999} for a pedagogical review). The temperature-dependent effective potential can be  obtained at one-loop order by adding a correction to the zero-temperature one-loop effective potential,
\be
V_{\textrm{eff}}(h,\f,T)=\Vtree(h,\f)+\Vone(h,\f)+\VT(h,\f,T).\label{eq:V-eff}
\ee
The finite-temperature correction is given by the following formula
\be
\VT(h,\f,T)=\frac{T^4}{2\pi^2}\sum_{a} n_a J_a\left(\frac{M_a(h,\f)^2}{T^2}\right),\label{eq:thermal-pot}
\ee
where the sum runs over particle species. $J_a$ denotes the thermal function, which is given by
\be
J_{F,B}(y^2) =  \int_0^\infty \dd{x} x^2 \log(1\pm e^{-\sqrt{x^2 + y^2}}),\label{eq:thermal-functions}
\ee
where the ``$+$'' sign is used for fermions ($J_F$), while the ``$-$'' for bosons ($J_B$). If one expands the bosonic thermal function in the high $T$ regime (small $M/T$, see appendix~\ref{app:high-T}) one can see that the finite-temperature correction can produce a term cubic in the fields, which can generate a potential barrier between the symmetric and symmetry-breaking minima leading to a strong first-order phase transition.

It has been shown that this basic formula for the effective potential is not enough since in the high-temperature limit higher loop contributions can grow as large as the tree-level and one-loop terms. This means that perturbative expansion in terms of  loops fails and one has to improve the computation scheme by resumming the class of problematic higher-order contributions. This is commonly attained by a resummation of the so-called daisy diagrams and shifting the field-dependent masses by inclusion of the high-temperature part of the self-energy corrections $M_a^2(h,\f)\to M_a^2(h,\f)+\Sigma$~\cite{Parwani:1992, Arnold:1992, Carrington:1992}. This approach alleviates the problem and provides reliable results in most of the cases. The formulas for the thermal masses of the gauge bosons and scalars can be found in appendix~\ref{app:thermal-masses}. There is no need for resummation of the fermionic contributions since they do not suffer from IR divergences~\cite{Espinosa:1992-2}.   

In the present paper we first use the approach described above to analyse the phase transition and gravitational wave signal in the SU(2)cSM. As we find that the nucleation temperature is fairly low, we aim to improve this approach by implementing a procedure that does not use the high-temperature expansion at any step and also includes subleading terms (see section~\ref{sec:improved-potential}).

\section{Physics of symmetry breaking\label{sec:symmetry-breaking}}

In this section we briefly explain the physical phenomena of electroweak phase transition and gravitational wave production.

\subsection{Radiative symmetry breaking\label{sec:RSB}}

In this work we are interested in models which possess classical conformal symmetry --- this means that the classical lagrangian cannot contain any dimensionful parameters. As a consequence, all the fields are massless and no energy scale exists at classical level. Moreover, the standard Brout-Englert-Higgs mechanism is not applicable due to the absence of the negative mass-term for the scalar field. Nonetheless, quantum effects can change the picture diametrically --- they can break conformal symmetry and generate particles' masses through the so-called dimensional transmutation also referred to as the Coleman-Weinberg (CW) mechanism~\cite{Coleman:1973} or RSB.

When models with extended scalar sector are considered, different complications with regard to searching for radiatively generated minima are introduced, from the scale dependence of the effective potential to the fact that  the effective potential becomes a function of more variables and consequently the vacuum structure of the potential can be rather complicated. Many simplified approaches to RSB have been developed, often constraining the effective potential to some direction in the field space. A recent extensive discussion of  RSB in models with extended scalar sectors can be found in ref.~\cite{Chataignier:2018RSB} (see also~\cite{Loebbert:2018}).  

In what follows to study the symmetry-breaking transition we  use the full one-loop effective potential being a function of two independent fields $(h,\f)$.  This allows us to study the transition between the minima in full generality. In particular, we can find the bubble profiles along the $\f$ and $h$ directions and determine the trajectory along which the field tunnels from the false to the true vacuum. For the sake of comparison we also employ the well-know approach of Gildener and Weinberg~\cite{Gildener:1976} with which the effective potential is reduced to a one-argument function. Here we briefly explain the method and a more extensive introduction is presented in appendix~\ref{app:Gildener-Weinberg}. The method is based on an observation that at a certain renormalisation scale, called here $\mgw$, the tree-level potential acquires a flat direction, along which it vanishes. It is then sufficient to study the effective one-loop potential along this direction as in the rest of the parameter space the tree-level potential dominates. The analysis is simplified radically by this assumption, however, also the applicability is limited. Using the Gildener-Weinberg approach one can only study transitions along a direction from the origin of the field space to the minimum which might not be the true direction of the transition. In the present paper we will verify whether one can obtain correct results for the GW spectra using the Gildener-Weinberg approach.

\subsection{Phase transitions}

The electroweak phase transition (EWPT), during which particles acquire mass, is believed to proceed through nucleation of bubbles of the ``true'' vacuum (state corresponding to the global minimum of the potential) in the sea of the  ``false'' vacuum (corresponding to a local minimum). At high temperature the Universe begins in the symmetric phase, where the ground state of the theory is located at the origin of the field space and the effective potential has no other minima. While the temperature decreases, another minimum forms and when it becomes the global minimum, tunnelling to this state becomes possible. The temperature at which the two minima are degenerate is referred to as the critical temperature, $T_c$. When a bubble  of true vacuum is created the potential energy inside it is lower than outside. However, there is also tension on the surface of the bubble which increases the energy of such a configuration. When the temperature is such that  large bubbles can form and the volume energy can win over the surface energy, nucleation of bubbles can proceed. The probability of creation of a bubble is described by the vacuum decay (nucleation) rate denoted by $\Gamma$. Additionally, in the cosmological context, the density of the bubbles is washed out by the expansion of the Universe and therefore the phase transition can be completed if at least 
 one bubble is formed per unit of time per Hubble volume. Temperature at which that happens is referred to as nucleation temperature, $T_n$.  The transition is completed when the bubbles percolate and we assume that this happens at the same temperature $T_n$. We will find that the nucleation temperature is much lower than the critical temperature, thus the energy of the false vacuum can in principle induce an inflationary period~\cite{Barreiro:1996}, we will not consider this scenario here, though (see section~\ref{sec:discussion} for discussion).

To find the nucleation rate and the bubble profiles one needs to find the so-called bounce solution and the three-dimensional Euclidean action of this solution, $S_3(T)$. This part of the analysis is relegated to an external programme AnyBubble~\cite{Masoumi:2017}.  Once the Euclidean action is known, the nucleation rate can be computed as\footnote{This formula assumes classical bubbles, better accuracy can be obtained by taking into account quantum effects in the exponential prefactor~\cite{Baacke:1993, Baacke:1995, Baacke:1995-2, Kripfganz:1994, Kripfganz:1995}, this is, however, beyond the scope of the present paper.}
$$
\Gamma\approx T^4 e^{-S_3(T)/T}.
$$

\subsection{Gravitational waves}

The GW remaining from the EWPT can be sourced by the  colliding bubbles of true vacuum as well as their interaction with the surrounding plasma. Therefore, certain properties of the bubbles are relevant for the computation of the GW signal. One of them is the velocity of the bubble wall $v_w$, which is expected to be close to 1 for the strong transitions that we will find~\cite{Bodeker:2009}. Nonetheless, the bubbles are not expected to run away~\cite{Bodeker:2017}, thus the dominant contribution to the GW signal is expected to come from 
the sound waves and turbulence in the plasma which are formed after collisions of the bubbles~\cite{Caprini:2015}. Since the turbulence is expected to be much less important than the sound waves, we limit the analysis here to the latter.

The vacuum decay rate $\Gamma$ is parametrised as
\be
\Gamma(t) \sim e^{-\beta t},\nonumber
\ee
from which it is clear that $\beta$ has an interpretation of an average inverse time scale of the transition~\cite{Apreda:2001}, which is another parameter relevant for the GW signal. The energy that is released during the transition (the latent heat), $\epsilon$ corresponds to the energy density difference between the true and the false vacuum. It consists of the difference in free energy (or effective potential) and additionally by an entropy variation. In the limit of large supercooling, $T_n\ll T_c$ the entropy contribution can be neglected~\cite{Marzola:2017}. For the gravitational wave signal it is crucial to determine the ratio of $\epsilon$ to the energy density of radiation at the time of the transition, $\rho_r^*$
\begin{align}
\alpha = \frac{\epsilon}{\rho_r^*},
\label{eq:alpha}
\end{align}
where the radiation density as a function of temperature is given by
\begin{align}
\rho_r(T) = \frac{g_* \pi^2}{30} T^4.\nonumber
\end{align}
We assume that the relativistic degrees of freedom $g_*$ are temperature independent during the time  of the transition.

The corresponding power spectrum of the GW signal is given by~\cite{Caprini:2015}
\begin{align}
h^2 \Omega(f) =2.65 \cdot 10^{-6} \qty(\frac{H_*}{\beta}) \qty(\frac{\kappa_v \alpha}{1+\alpha})^2 \qty(\frac{100}{g_*})^{1/3} v_w S_\text{sw}(f),
\label{eq:GWspectrum}
\end{align}
where the spectral shape is defined by
\begin{align}
S_\text{sw}(f) = (f/f_\text{sw})^{3}\qty(\frac{7}{4+3 (f/f_\text{sw})^{2}})^{7/2},
\label{eq:spectral}
\end{align}
and the peak frequency today is
\begin{align}
f_\text{sw}= 1.9 \cdot 10^{-2} \text{mHz} \frac{1}{v_w}  \qty(\frac{\beta}{H_*}) \qty(\frac{T_*}{100 \text{GeV}}) \qty(\frac{g_*}{100})^{1/6}.
\label{eq:peakf}
\end{align}
The parameter $\kappa_v = \rho_v/\rho_\text{vac}$ is defined as the fraction of vacuum energy that gets converted into bulk motion of the fluid~\cite{Caprini:2015}. Interestingly, in the case of large supercooling $\alpha  \gg 1$ the dependence of the spectrum on $\alpha$ drops out.

\section{Phase transition and resulting gravitational waves in SU(2)cSM\label{sec:results}}

\subsection{Numerical procedure}
Our aim is to assess the GW signal that arises from the phase transition in the early Universe within the framework of SU(2)cSM. We perform the computations for the benchmark points shown in Table \ref{tab:benchmark}, defined such as to probe the parameter space where the potential is stable up to the Planck scale (following the analysis of ref.~\cite{Chataignier:2018RSB}). They are characterised by the couplings $\lambda_1$, $\lambda_2$, $\lambda_3$ and $g_X$, at the scale $\mu=v=246\g$. The VEV of the SM scalar field is equal to $v=246\g$, while we denote the VEV of the new scalar $\f$ by $w$ (both computed at $\mu=v$). The Higgs one-loop mass is fixed to $125\g$. The cosine of the mixing angle that diagonalises the mass matrix and simultaneously rescales (with respect to the SM) the couplings of the Higgs boson to the gauge bosons, is not within the experimental bounds for all the points, since we aimed at inspecting the parameter space more broadly. The points that satisfy all the constraints are the points number 3, 4, 6 and 7 (see the discussion in ref.~\cite{Chataignier:2018RSB}, following refs.~\cite{Robens:2016, Robens:2015, Ilnicka:2018}). We take these points as representative points in the parameter space to find out whether SU(2)cSM can produce an observable spectrum of GWs. Furthermore, we use the benchmark point with label 7 to illustrate our findings in more detail. 
\renewcommand{\arraystretch}{1.3}
\setlength{\tabcolsep}{7pt}
\begin{table}[ht]
\begin{center}
\begin{tabular}{ccccccccccc}
\# & $\lambda_1$ & $\lambda_2$ & $\lambda_3$ & $g_X$ & $w$ [GeV] & $T_c$ [GeV]  & $T_n$ [GeV]  & $\beta/H_*$ & $\alpha$ \\
\hline
1     & 0.175       & -0.0049     & -0.0038     & 0.83  & 2200            &  281   &    22      & 449  &  597        \\
2     & 0.149       & -0.0065     & -0.0058     & 0.94  & 1774            &  256   &   27 &  323   & 213            \\
3     & 0.119       & -0.0013     & -0.0136     & 1.01  & 3611            &  568   &  64 &  238  &  131           \\
4     & 0.122       & -0.0050    & -0.0104     & 1.05  & 1860            &  302   &        34 &300    &  137           \\
5     & 0.166       & -0.0083     & -0.0063     & 0.97  & 1648            &  244   &    25 &  327  &  210 \\
6     & 0.120       & -0.0019     & -0.0079     & 0.92  & 2991            &  428   & 39 &   419 &    345 \\
7     & 0.124       & -0.0030     & -0.0047     & 0.85  & 2411            &  318   &    28 &  434  &       361 \\ 
8     & 0.139       & -0.0095     & -0.0093     & 1.08  & 1426            &  236   &    29 &   250 &     87
\end{tabular}
\label{tab:benchmark}
\caption{Parameters and characteristics of the phase transition for the analysed benchmark points.}
\end{center}
\end{table}

The numerical procedure to obtain the GW signal from the finite-temperature effective potential for a given benchmark point is structured as follows:

\begin{enumerate}
\item We compute the finite-temperature effective potential on a grid in the $(h,\varphi)$ plane. To account for reliable results at all temperatures, we use for the thermal functions $J_{F, B}$ the numerical solution to the integral \eqref{eq:thermal-functions} with the aid of the publicly available numerical implementation for the thermal functions of ref. \cite{Fowlie:2018}. We interpolate from the grid to yield the finite-temperature effective potential. This procedure is repeated for different temperatures with a discrete temperature increment.

\item Next, the potential is minimised  to find all local and global minima and the critical temperature. Knowing the true vacuum, false vacuum and the effective potential at a given temperature, we can compute the bubble profiles in the two-field case and the corresponding euclidean action $S_3(T)$. These are found with the use of a publicly available Mathematica package AnyBubble~\cite{Masoumi:2017}.\footnote{It should be underlined that we use the full effective potential, including both of the scalar fields. The price that is paid for this generality is that the code exhibits certain numerical instability, returning no results for certain values of temperature.}

\item We assume a radiation dominated Universe (see section~\ref{sec:discussion} for discussion),  the condition that at least one bubble nucleates per Hubble horizon translates then into
\be
\int_{T_n}^{T_c} \frac{\dd T}{T} \qty(\frac{2 \xi M_\text{Pl}}{T})^4 e^{-S_3(T)/T} \sim 1, \label{eq:nucleation-temp}
\ee
where $\xi = \frac{1}{4 \pi} \sqrt{\frac{45}{\pi g_*}}$ and $M_{\textrm{Pl}}$ is the Planck mass, $M_{\textrm{Pl}}=1.22\cdot 10^{19}\g$ (see ref.~\cite{Quiros:1999} for a derivation). We use this equation to determine the nucleation temperature $T_n$. To compute the time scale we use the formula \cite{Apreda:2001}
\be
\frac{\beta}{H_*} = T_n \eval{ \frac{\dd(S_3/T)}{\dd T}}_{T=T_n},\label{eq:beta}
\ee
by using the slope of a linear fit to the values of $S_3(T)/T$ around the nucleation temperature. We should note, that the error of this fit is high due to fluctuations of the values of $S_3(T)/T$ computed  by AnyBubble. Therefore, the obtained values for $\beta/H_*$ are afflicted with considerable numerical error. More details on how $\beta$ is computed can be found in appendix~\ref{app:beta}. The parameter $\alpha$ is computed with the use of eq.~\eqref{eq:alpha}. 
\item Finally, we obtain the SGWB by eqs. \eqref{eq:GWspectrum}, \eqref{eq:spectral} and \eqref{eq:peakf}, where we assume $v_w = 1$, which is justified since we deal with very strong transitions,  and $\kappa_v=1$, which is a good approximation, since the parameter $\alpha$ that we find is large (for more detailed formula see ref.~\cite{Caprini:2015} and references therein).
\end{enumerate}

\subsection{Nucleation of bubbles\label{sec:bubbles}}

When models with extended scalar sector are under scrutiny, the issue of the nucleation of bubbles becomes more complicated. The questions we address in the present section are as follows --- how does the nucleation proceed? Are nonzero VEVs for both of the fields produced simultaneously or is it a sequential process? What is the critical temperature? What is the nucleation temperature? In this section we focus on a single benchmark point (number 7 from table~\ref{tab:benchmark}) to discuss in detail the dynamics of the phase transition and later we comment on the other choices of parameters.

\begin{figure}[t]
\center
\includegraphics[height=.28\textwidth]{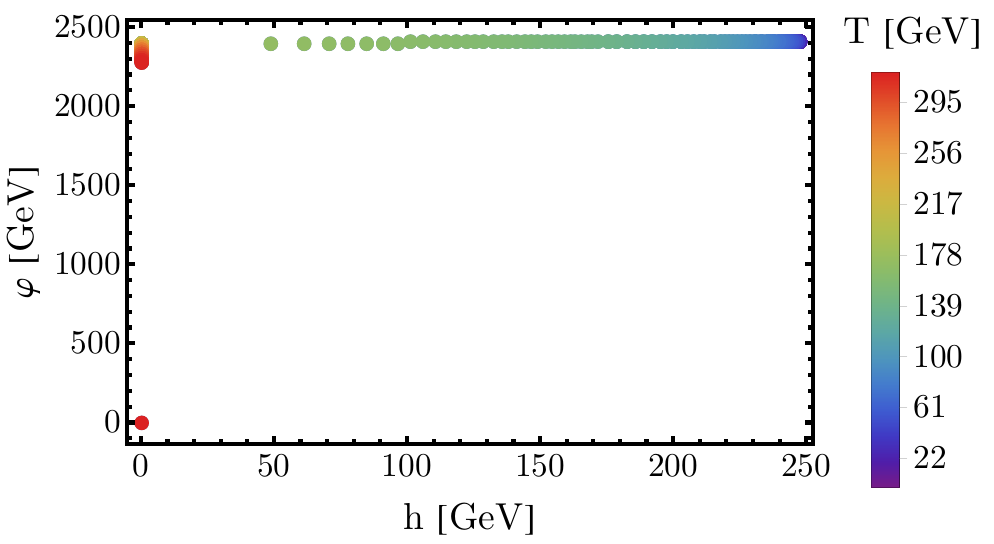}
\includegraphics[height=.28\textwidth]{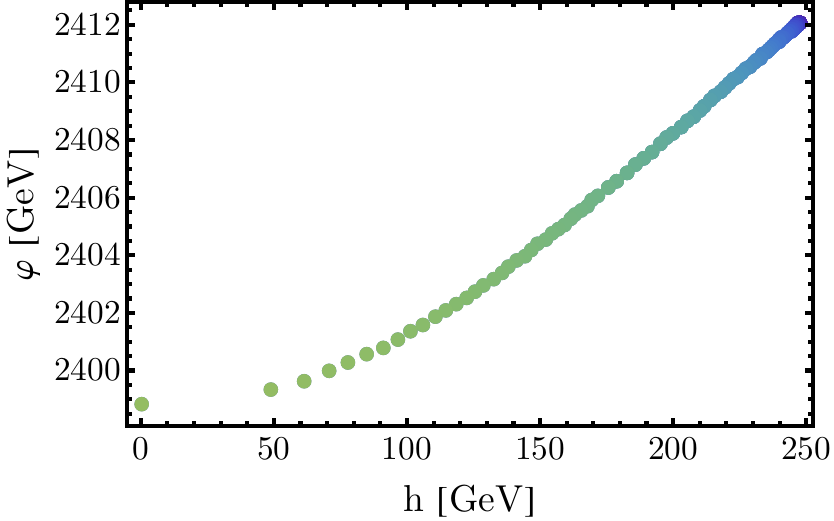}
\caption{The trajectory of the global minimum of the effective potential in the $(h,\f)$ plane. The colours of the points indicate the temperature to which a given point corresponds. The plot in the right panel shows with better resolution the trajectory of the minimum after nonzero VEV for $\f$ is acquired. The colour coding is common for both of the plots.\label{fig:location-of-min}}
\end{figure}

Let us start from analysing how the location of the global minimum changes with the temperature. Figure~\ref{fig:location-of-min} shows the location of the global minimum in the $(h,\f)$ plane, the temperature to which a given point corresponds is encoded in the colour of the point. At high temperature we start from a symmetric minimum at the origin of the field space, as the temperature decreases below the critical temperature which in this case is at $T_c=318\g$, the symmetric minimum is reduced to a local one and a new global minimum at nonvanishing $\f$ (but $h=0$) arises. It stays like this for some time and later a nonzero VEV for the SM scalar field is developed, later gradually reaching 246$\g$ at zero temperature. This shows that the tunnelling can proceed either between the symmetric vacuum and a vacuum where only $\f$ acquires a VEV or directly to the state where both of the scalars have nontrivial VEVs. To answer this question one needs to find the nucleation temperature and the bubble profile.

\begin{figure}[t]
\center
\includegraphics[width=.48\textwidth]{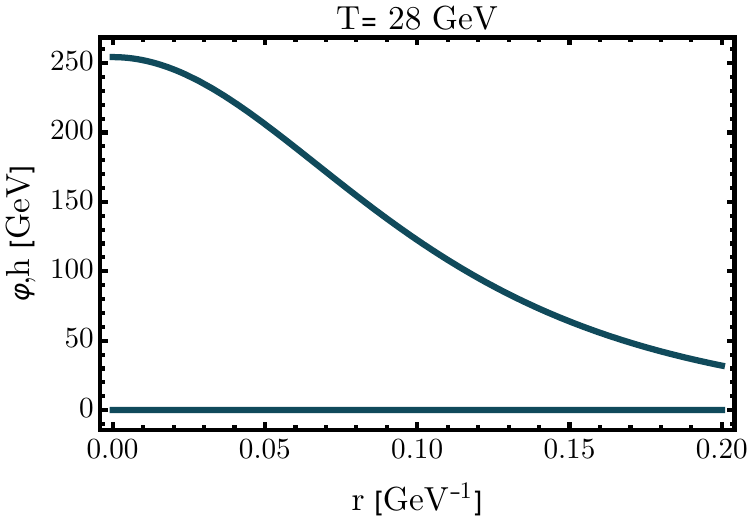}
\caption{The bubble profile at the nucleation temperature.\label{fig:bubble}}
\end{figure}

\begin{figure}[t]
\center
\includegraphics[height=.4\textwidth]{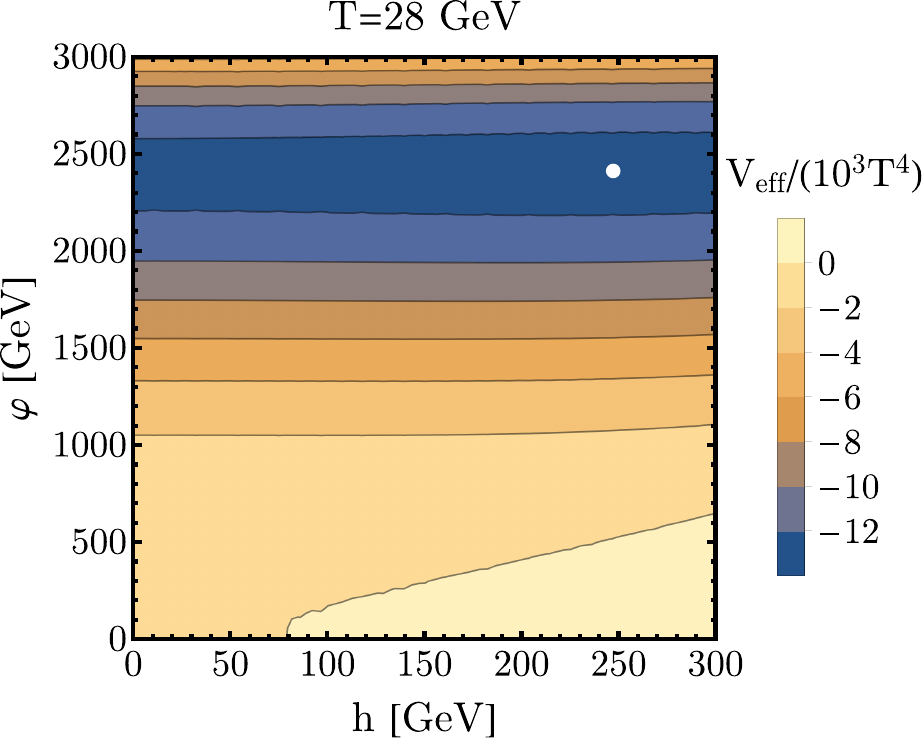}
\includegraphics[height=.4\textwidth]{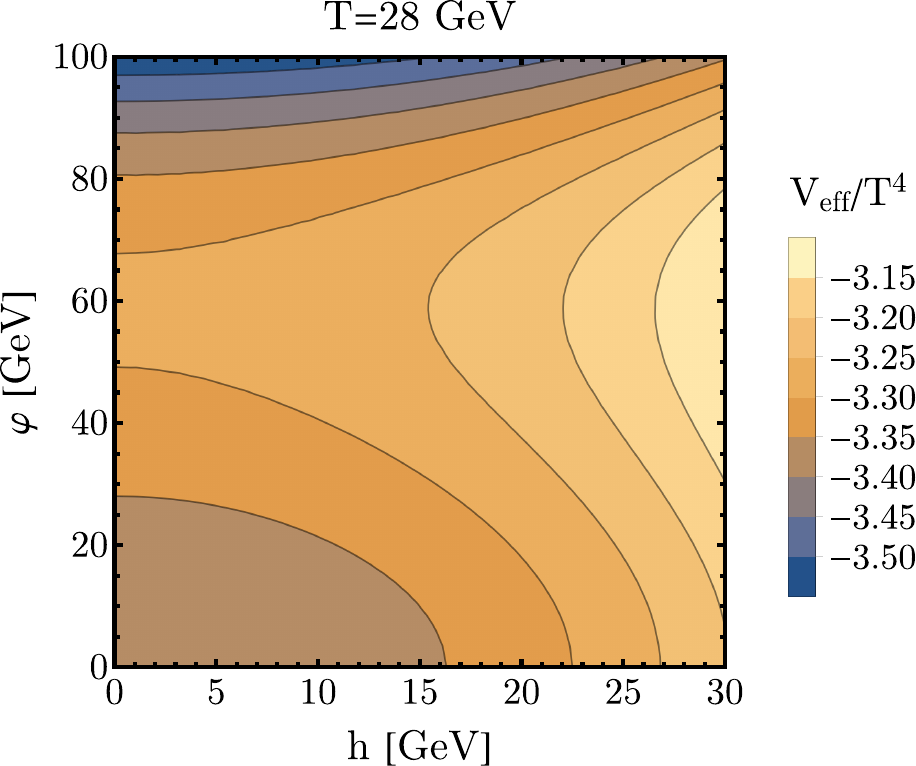}
\caption{The contour plot of the effective potential at the nucleation temperature. In the left panel the true minimum is marked with the white dot, in the right panel the region around the symmetric minimum is shown, capturing the saddle point. Note that the colour coding is different in the two panels. \label{fig:contours-potential}}
\end{figure}

The nucleation temperature for the discussed benchmark  point is found to be equal to $T_n=28\g$,  which is much below the critical temperature. This suggests what we have anticipated --- large degree of supercooling and possibly very strong transition. The bubble profile for the benchmark point that we study  is shown in figure~\ref{fig:bubble}.  First important point to note is that the tunnelling takes place only in the $\f$ direction. At the nucleation temperature, the global minimum is already located at a point with nonvanishing $v$ and $w$, therefore after emerging from under the barrier, at $\f\approx 255\g$ and $h=0$, the field rolls down the potential to reach the global minimum. The value of $\f$ at which the field arrives after the tunnelling  is much lower than the value of the VEV, which simply means that it is more energetically (in terms of the free energy) favourable to tunnel to a point which is not at the minimum of the potential and subsequently roll down to the real minimum. In figure~\ref{fig:contours-potential} the contour plot of the potential at the nucleation temperature is showed, the plot in the left panel captures both the global minimum (marked with the white dot) and the local minimum at the origin of the field space. The right panel shows small region around the origin, displaying the local minimum and the saddle point at (or very close to) the $\f$ axis. It can be thus understood that the tunnelling trajectory passes through the saddle point, where the potential barrier is the thinest. Moreover, the global minimum is very shallow along the $h$ direction, which tells us that not much energy can be gained by tunnelling directly to the global minimum. In fact, apparently this gain would be less than the energy consumed to penetrate the deeper potential barrier in the direction to the global minimum, which is a possible explanation of the aforementioned behaviour.

It is important to note that we would not be able to reach the conclusions about the tunnelling trajectory presented above if we had not used the full two-field effective potential. As we noted in section~\ref{sec:RSB}, the Gildener-Weinberg approach is frequently employed to study symmetry breaking in models with classical conformal symmetry. With this method one studies the effective potential along the direction between the origin of the field space and the minimum. Therefore, using the Gildener-Weinberg method it is not possible to reproduce the tunnelling pattern presented in figure~\ref{fig:bubble}. On the other hand, it is consistent with the sequential approach, where the symmetry breaking is considered independently of the SM sector (see ref.~\cite{Chataignier:2018RSB} for a discussion of the zero-temperature case).

Let us finalise this section by commenting on the other benchmark points presented in table~\ref{tab:benchmark}. As can be seen, the critical temperatures vary between 236 and 568$\g$, where higher zero-temperature VEVs of the scalar field correlate with higher critical temperatures and higher nucleation temperatures. The nucleation temperatures vary between 22 and 64$\g$, therefore, all the points display large supercooling and strong phase transition. Moreover, for all the points the trajectories of the minimum parametrised by temperature and the bubble profiles resemble those presented in figures~\ref{fig:location-of-min} and \ref{fig:bubble}. Therefore, we can conclude that the scenario described above is representative for the whole parameter space of interest.

\subsection{Gravitational wave signal}

\begin{figure}[t]
\center
\includegraphics[height=.35\textwidth]{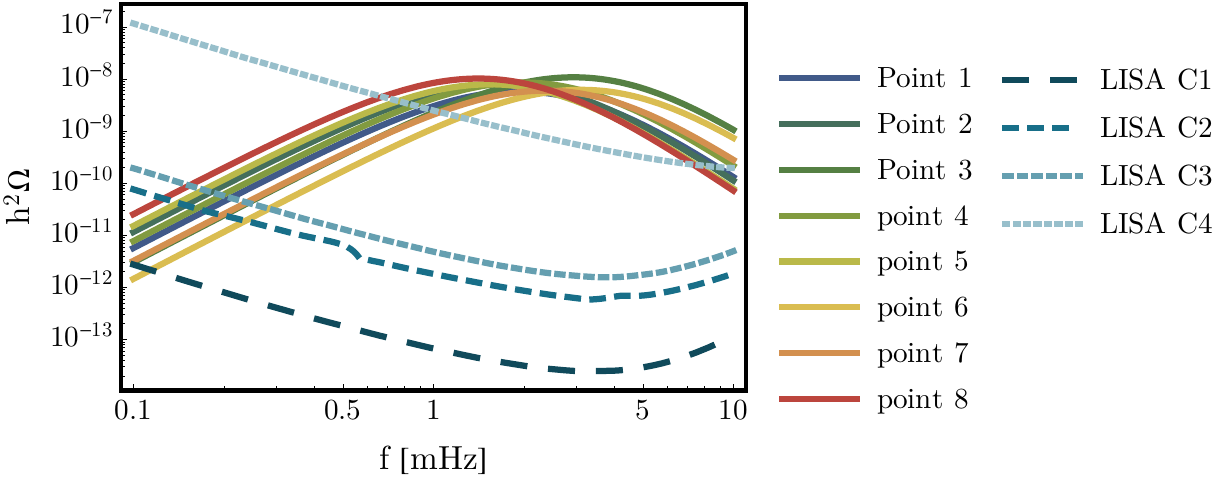}
\caption{Gravitational wave signals for the benchmark points analysed (solid lines), LISA sensitivity curves corresponding to different configurations (dashed lines).\label{fig:GW}}
\end{figure}

Knowing the bubble profiles we are able to compute the parameters needed to find the GW spectra. As should be clear from table~\ref{tab:benchmark}, typically the transition proceeds after large supercooling, the latent heat release is significant and the transition proceeds fast (large $\beta/H_*)$ , thus we expect  a very strong transition and hence strong GW signal. The actual results for the benchmark points that we analysed are presented in figure~\ref{fig:GW}, together with the LISA sensitivity curves corresponding to different configurations~\cite{Caprini:2015}. Indeed the GW signals predicted are extremely strong, which makes the SU(2)cSM an interesting model from the point of view of future GW astronomy. The model will be clearly testable in the GW experiments.

\begin{figure}[t]
\center
\includegraphics[height=.35\textwidth]{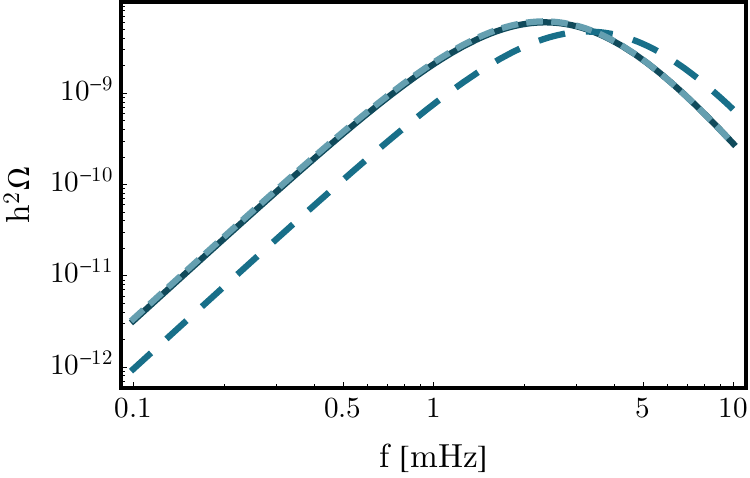}
\caption{Comparison of the GW signals predicted with the use of the full two-field potential (solid line), the Gildener-Weinberg method (long-dashed line) and the potential along the $h=0$ direction (short-dashed line, overlaps the solid line).\label{fig:GW-GW}}
\end{figure}

In section~\ref{sec:bubbles} we have explained that the tunnelling proceeds along the $\f$ direction in the field space and we noted that because of that the Gildener-Weinberg method of studying symmetry breaking cannot reproduce the correct picture of bubble nucleation since it only takes account of the potential along one direction in the field space, the direction from the origin of the field space to the minimum. However, this direction turns out to be close to the $h=0$ direction, since the VEV of $\f$ is much greater than the VEV of $h$ (see table~\ref{tab:benchmark}). Therefore, it is possible that the Gildener-Weinberg method gives approximately correct results for the gravitational wave spectra. To verify whether that is indeed the case we computed the GW signal predicted by the Gildener-Weinberg potential. Figure~\ref{fig:GW-GW} shows the predicted GW signal for benchmark point 7 computed from the full two-field potential (solid line) and using the Gildener-Weinberg approach (long-dashed line). Moreover, we present the result that we would obtain by considering the potential along the $\f$ direction only (setting $h=0$), which in the light of the shapes that we found for the bubble profiles seems to be a reasonable approximation.\footnote{Note, however, that this might not be the case in other regions of the parameter space or in different models.} One can see that the full two-dimensional and the $h=0$ approaches agree very well. The Gildener-Weinberg results, however, show certain discrepancy both in the peak frequency and in the intensity of the signal.  However, as we have mentioned before (and we explain in detail in appendix~\ref{app:beta}) in our study there are considerable uncertainties in the evaluation of $\beta$, which could partially account for the inaccuracy of the Gildener-Weinberg method or make the discrepancy between $h=0$ approach and multi-field approach larger.

\subsection{Discussion\label{sec:discussion}}

At the final stage of preparation of the present paper several interesting articles related to the issue of gravitational wave production during first order phase transition appeared. Ref.~\cite{Baldes:2018} presents an interesting analysis of the GW spectra within the SU(2)cSM also taking into account dark matter relic abundance constraints, assuming different production mechanisms and includes the effect of astrophysical foregrounds. The results confirm that the GW signal is expected to be very strong and observable by LISA. Note that our study supplies evidence for the assumption made in ref.~\cite{Baldes:2018} that it is enough to study the phase transition along the $\f$ direction only. Ref.~\cite{Kanemura:2018} discusses the possibility of constraining the fundamental parameters of theoretical models by measurements of gravitational wave signals complementarily to accelerator experiments (see also refs.~\cite{Figueroa:2018, Alves:2018} on similar topic), also within models with classical scale invariance. This possibility also shows the necessity of pursuing the effort of providing reliable theoretical predictions. Finally, ref.~\cite{Ellis:2018} discusses the role of false-vacuum-energy-induced inflation, relevant for the cases with large supercooling, for the possibility of completing the phase transition. In the context of classically conformal models it is noted that if a period of inflation is present, the transition can still be successfully completed and moreover, the plasma can be diluted sufficiently for the bubble-wall collisions to be an effective source of GW. However, this requires very low nucleation temperature, $T_n\lesssim 10$\ MeV, which we do not find in our case. We have checked for our benchmark point the influence of including the vacuum energy, in addition to the radiation energy, in eq.~\eqref{eq:nucleation-temp} for the nucleation temperature. We have found a change in $T_n$ around $0.5\g$ and the resulting changes in $\beta_*/H$ and $\alpha$ within our numerical errors. Thus, implementing the results of ref.~\cite{Ellis:2018} would not change our results significantly. In ref.~\cite{Ellis:2018} it is also underlined that the results of numerical simulations on which formulas used to evaluate GW spectrum from sound waves are based, require the source for gravitational waves to last sufficiently long to be reliable. It is not the case in the models studied in ref.~\cite{Ellis:2018} and this may also apply to the present analysis.

It is in order to note that the results that we presented for the GW spectra are also afflicted with other difficulties. First of all, there are numerical issues related to finding the bubble profiles and their properties discussed in appendix~\ref{app:beta}. To some extend they are the price to be paid for not constraining the field space to one dimension. In the cases where one verifies, as we did above for the benchmark points, that the transition only proceeds along one direction, a possibility of improving the accuracy is to perform the computations with the one-dimensional potential, as was done in ref.~\cite{Baldes:2018}. Furthermore, we do not discuss the issue of renormalisation group improvement of the effective potential (see e.g.\ refs~\cite{Chataignier:2018, Chataignier:2018RSB} and references therein) since we focus on the thermal effects. Nonetheless, it can be important in the regimes where the zero-temperature part of the effective potential dominates. Another issue is the gauge dependence of the results obtained from the effective potential~\cite{Wainwright:2011GW, Chiang:2017}. Last but not least, the dynamics of bubble formation and heat transfer in the plasma is not fully understood yet. There have been many developments in this field over the recent years (see e.g.~\cite{Bodeker:2017, Ellis:2018}), which change our understanding of these processes.

Being aware of all these issues, in the following part of the paper we focus on improving our understanding 
of the early stages of the phase transition in conformal extensions of the standard model. As it turns out, a typical transition 
proceeds after a large amount of supercooling, resulting in a very strongly first order transition. In what follows we propose a thermal resummation technique based on the gap equation which should be reliable in the full range of temperatures and provides better accuracy by including also subleading terms as a first step in the refined treatment of phase transitions.

\section{Handling the effective potential\label{sec:improved-potential}}

As explained above, the thermal effective potential requires resummation of higher-order diagrams in order to be reliable. This is usually completed by resummation of the leading daisy diagrams, which leads to an improved propagator. In order to read-off the corrections to masses --- the thermal masses --- the high-temperature expansion is performed (an expansion in $M/T$, where $M$ is a zero-temperature mass). There are arguments showing that the thermal masses should be valid beyond the high-temperature expansion~\cite{Carrington:1992}. As the thermal masses are proportional to $\alpha T^2$, where $\alpha$ is the relevant coupling, at low temperature they give a small contribution, as expected. At temperatures where $M  \sim T$ the thermal correction scales as $\alpha T^2 \sim \alpha M^2$, i.e.\ is suppressed by the coupling constant with respect to the zero-temperature contribution. These arguments are sufficient in most of the cases. However, the results for intermediate and low temperatures, while being suppressed, can still not be reliable (even though they introduce a small error in the final result). While the temperatures relevant for the phase transition analysed in the present paper cover a wide range, since the nucleation temperature is approximately an order of magnitude below the critical temperature, we need results that are  fully valid in the intermediate region. Moreover, the gauge coupling of the SU(2)$_X$ group is rather large, which weakens the suppression argument presented above. Therefore, in this section we develop a formalism, based on the 2PI effective action approach, which allows to resum thermal corrections to masses without referring to the high-temperature expansion. Its additional advantage is that it resums subleading contributions, enhancing the accuracy of the results.

\subsection{Gap-improved effective potential\label{sec:gap}}

As a means of performing resummations in a consistent way, Cornwall, Jackiw and Tomboulis generalised the 1PI effective action approach to an $n$PI approach \cite{Cornwall:1974} (for a pedagogical introduction see ref.~\cite{Berges:2004}).  The $n$PI effective action has functional dependence not only on the one-point function, as it is in the 1PI approach used earlier in this paper, but also on higher $n$-point functions, e.g.\ the 2PI effective action depends on the field $\f$ and the resummed propagator $G$.  Having both $\f$ and $G$ one can probe the configuration space better than with the use of the usual 1PI formalism. At one-loop level the 2PI and 1PI actions are exactly equivalent~\cite{Berges:2004}. Therefore, one needs to go to two-loop order to fully appreciate the benefits of the 2PI approach. In the present paper we aim at improving the accuracy and reliability of finite-temperature computations while not  magnifying the computational effort too much. Therefore, we will use a simplified approach, which amounts to improving the one-loop effective potential using 2PI tools. The computation of a full 2PI effective potential (see e.g.\ ref.~\cite{AmelinoCamelia:1992, AmelinoCamelia:1993, Pilaftsis:2015, Pilaftsis:2017}) would allow for exact consistency of the computations, however, we leave it for future work.

The method we propose amounts to computing the one-loop effective potential using an improved propagator, being a solution of the gap equation. It is a generalisation of the thermal-mass approach, where the shift in masses can also be interpreted as improving the propagator. By solving the gap equation we improve the propagator self-consistently, including not only the leading but also the subleading terms.

The gap equation for the inverse propagator can be written in terms of masses if one assumes an ansatz for the improved propagator $G$,
\begin{align}
G(k) = \frac{i}{k^2-M^2(h,\varphi, k)}\nonumber
\end{align}
as
\begin{align}
M^2(h,\varphi, k) = m^2(h,\varphi) + \Sigma(h, \varphi, M,  k),\notag
\end{align}
where $m$ represents the unimproved mass and the self-energy is evaluated using the improved propagator. It is important to note that the equation is self-consistent, following the 2PI approach, i.e.\ the self-energy is evaluated using the improved propagator. This means that the improved propagator that we obtain includes a wider range of diagrams than the standard daisy-improved propagator. With this method we are able to resum also subleading sunset diagrams, as well as iterated loops. 
 Moreover, we do not rely on the high-temperature expansion of the self-energy.

In principle, we should write such a gap equation for each particle in the theory, in particular for the particles that are known to spoil perturbativity of the thermal effective potential, i.e.\ the longitudinal components of the gauge bosons and the scalars. These gap equations would be all coupled (because different particles contribute to the self-energy) leading to a large complexity of the problem. In this paper, as a first step towards introducing a more reliable formalism for studying phase transitions at low temperatures, we focus on improving the scalar sector. Thus, we solve the set of coupled gap equations for the scalars, while using thermal masses of the gauge bosons in the contributing diagrams. Lacking the full solution, we introduce a patching function which turns on and off the thermal-mass correction for the $X$ boson, as its coupling is rather large and thus the high-temperature results are not automatically suppressed by the coupling (see appendix~\ref{app:thermal-masses}). It is important to note, that we do not rely anyhow on the high-temperature expansion in the scalar contributions. A next step, that is beyond the scope of the present paper, would be to implement such a procedure also for the gauge bosons, which would completely remove the high-temperature assumption from the analysis. This would be an advance especially for models where the phase transition proceeds after large super-coolings, which is the case in the SU(2)cSM model. 

Another simplification that we assume is to consider the zero-momentum limit of the gap equation. This allows us to use the effective potential to compute the self-energy contribution and also removes the momentum dependence from the improved mass. Having a momentum-independent improved mass we can compute the one-loop potential in the traditional way, obtaining as a result the well-known form of the effective potential with the masses replaced by the improved ones. This simplifies our analysis significantly. This assumption has been studied in the literature~\cite{AmelinoCamelia:1993} and one should keep in mind that improvements in this direction are in order.

Applying the approach described above we can write down the set of gap equations for the scalar sector of the SU(2)cSM as follows
\begin{align*}
M_h^2(h,\varphi, T) &= m_h^2(h,\varphi)+ \eval{\pdv[2]{}{h}\qty( V^{(1)}(h,\varphi) + V^\text{T}(h,\varphi)) }_{m_h\to M_h, m_\varphi \to  M_\varphi, m_{h \varphi} \to  M_{h\varphi}},\\
M_\varphi^2 (h,\varphi, T)&= m_\varphi^2(h,\varphi)+ \eval{\pdv[2]{}{\varphi}\qty( V^{(1)}(h,\varphi) + V^\text{T}(h,\varphi)) }_{m_h\to M_h, m_\varphi \to  M_\varphi, m_{h \varphi} \to  M_{h \varphi}} , \\
M_{h\varphi}^2 (h,\varphi, T)&= m_{h\varphi}^2(h,\varphi) + \eval{\pdv[2]{}{h}{\varphi}\qty( V^{(1)}(h,\varphi) + V^\text{T}(h,\varphi)) }_{m_h\to M_h, m_\varphi \to  M_\varphi, m_{h \varphi} \to  M_{h \varphi}},
\end{align*}
where $m_h$, $m_{\f}$ and $m_{h\f}$ stand for the elements of the tree-level mass matrix defined in eq.~\eqref{eq:tree-level-mass-matrix}. As explained above, the scalar improved masses denoted by capital $M$ are the variables we solve for and therefore we do not assume, e.g.\ standard thermal corrections for them. The gauge bosons contribute with their thermally-corrected masses. In the thermal term, $\VT$, we use the full thermal functions as given in eq.~\eqref{eq:thermal-functions} (using the implementation of ref.~\cite{Fowlie:2018}). Solving these equations by iterations yields the improved mass matrix
\begin{align*}
M^2 (h,\varphi,T) = \begin{pmatrix}
M_h^2 & M_{h \varphi}^2 \\ M_{h \varphi}^2 & M_\varphi^2 
\end{pmatrix}.
\end{align*}
We then compute the improved effective potential, denoted as $V$, using the standard formulas~\eqref{eq:one-loop}, \eqref{eq:V-eff}, \eqref{eq:thermal-pot} but with scalar masses in the thermal part substituted by the eigenvalues of the improved mass matrix.

\subsection{Impact of the improvement}

\begin{figure}[t]
\center
\includegraphics[height=.32\textwidth]{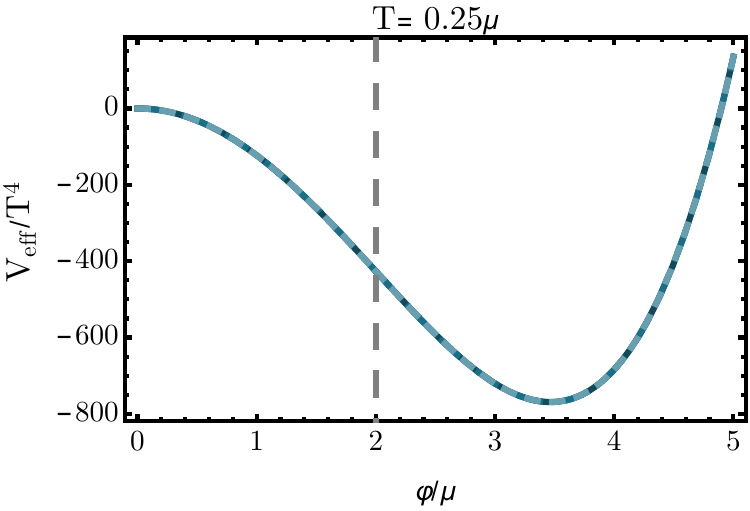}\hspace{.2cm}
\includegraphics[height=.32\textwidth]{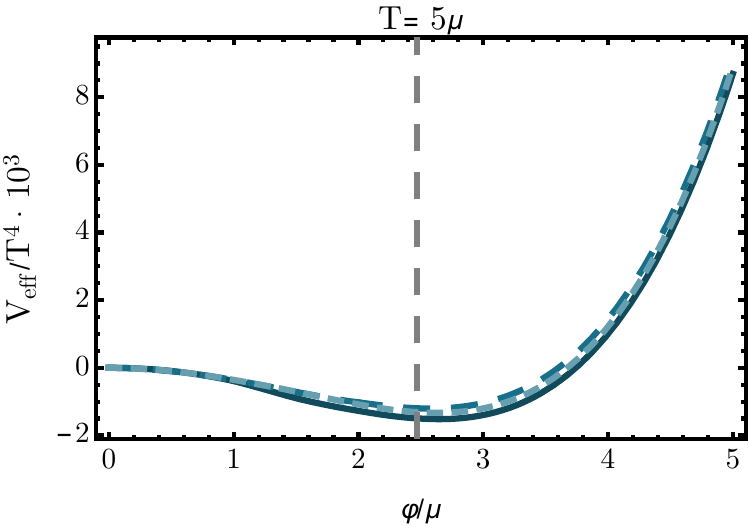}\\
\includegraphics[height=.32\textwidth]{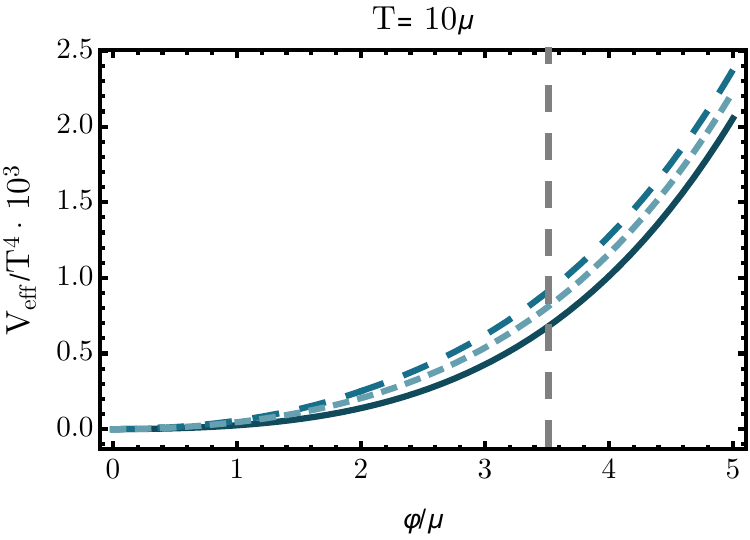}\hspace{.2cm}
\includegraphics[height=.32\textwidth]{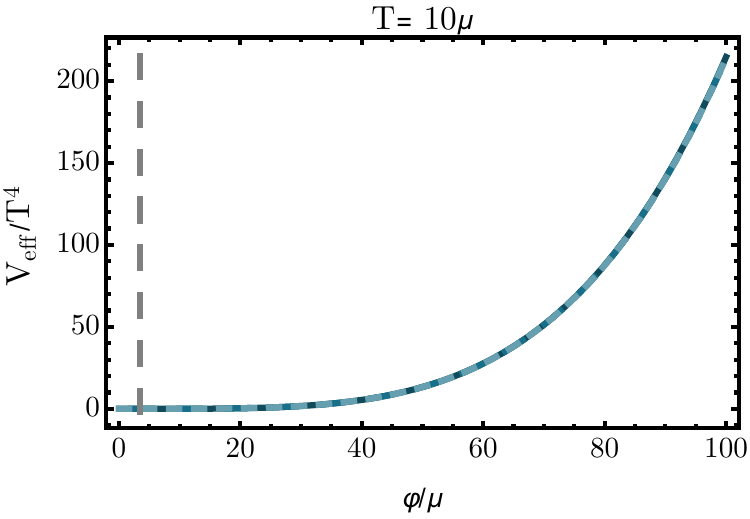}
\caption{Comparison between different approximations to the effective potential in pure scalar theory: the improved effective potential (solid line), thermal potential with daisy resummation (short-dashed line), thermal potential without daisy resummation (long-dashed line). The parameters of the potential are fixed to $m_0^2=-\mu^2$, $\lambda=0.5$, different plots correspond to varying temperature, low ($T=0.25\mu$), intermediate ($T=5\mu$, $T=5\mu$) and high ($T=10\mu$). Dashed lines indicate where the zero-temperature mass is equal to the thermal mass (to the left from the line the zero-$T$ mass is below the thermal mass).\label{fig:improved-pot-scalar-model}}
\end{figure}

Let us discuss the impact of the improvement procedure described in section~\ref{sec:improved-potential} on the effective potential. Since the resummation we perform is limited to the scalar sector we start from comparing the improved and unimproved potentials in a simple scalar model. The simplest conformally symmetric model is the pure scalar theory, $\lambda\f^4$, however it does not feature RSB~\cite{Coleman:1973}. Therefore, as a toy model, we consider the massive $\lambda\f^4$ model with the following classical potential
\begin{align*}
V^{(0)}(\varphi) = \frac{m_0^2}{2} \varphi^2 + \frac{\lambda}{4!} \varphi^4.
\end{align*}
The dimensionful parameters of this theory are the mass parameter $m_0$ and the temperature $T$. We quantify these parameters in the units of the renormalisation scale $\mu$, in practice setting $\mu$ to one in our computations. The results are visualised in figure~\ref{fig:improved-pot-scalar-model}. We compare the improved potential, the standard thermal potential including the daisy resummation and the thermal potential without daisy resummation. By the dashed line we indicate where the zero-temperature mass is equal to the thermal mass. To the left from that line the thermal corrections are most important but also the thermal-mass-approximation should work fine. Around the line our method of resummation is more reliable and potentially can improve the results. At very low temperatures (upper left panel) all three potentials coincide since they reduce to the zero-temperature result.  At higher temperatures (upper right and lower left panels), the improved potential starts to differ from the remaining two, which was to be expected. At high temperatures, but taking large field limit (lower right panel) the three potentials are indistinguishable since in this limit again the zero-temperature part dominates, since the field-dependent mass is larger than the temperature. This simple example confirms that the resummation procedure that we propose has some impact on the potential at intermediate temperatures, however, the impact is not very large. 

\begin{figure}[t]
\center
\includegraphics[height=.32\textwidth]{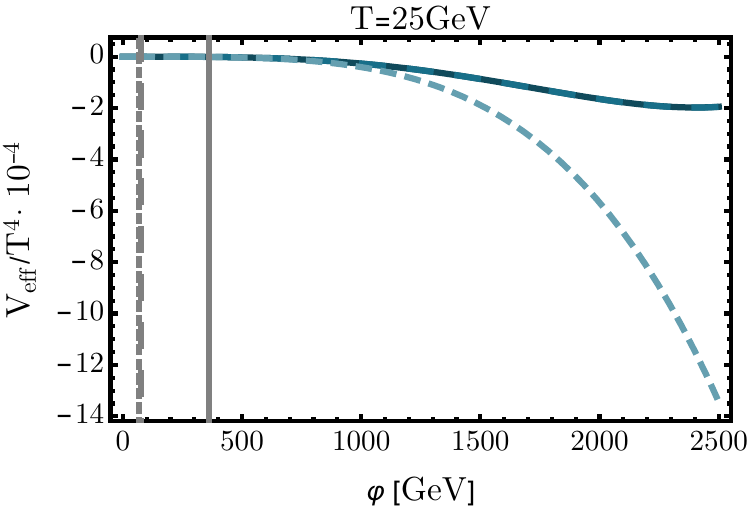}
\includegraphics[height=.32\textwidth]{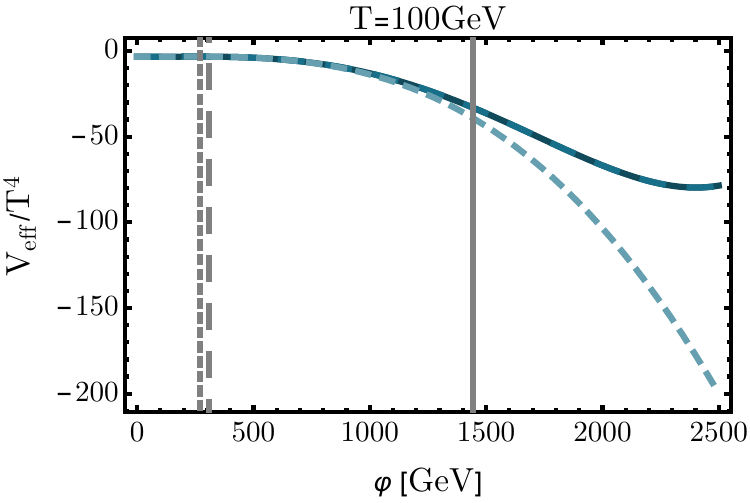}\\
\includegraphics[height=.32\textwidth]{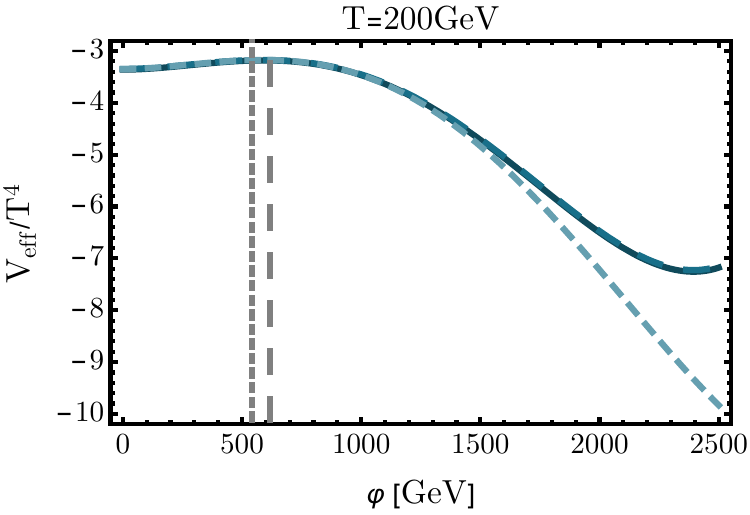}
\includegraphics[height=.32\textwidth]{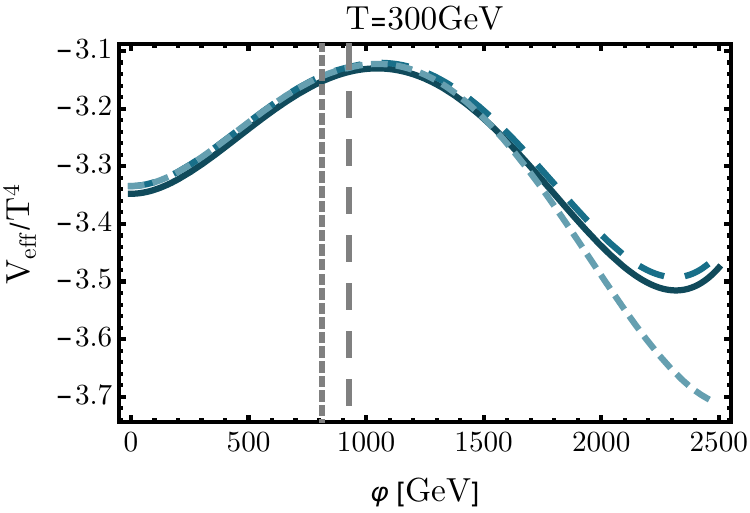}
\caption{Comparison between different approximations to the effective potential of the SU(2)cSM for the benchmark point number 7: improved effective potential (solid line), thermal effective potential with daisy resummation (long-dashed line) and thermal effective potential with daisy resummation and high-temperature expansion of the thermal functions (short-dashed line). Different plots correspond to varying temperature, low ($T=25\g$), intermediate ($T=100\g$, $200\g$) and high ($T=300\g$). Vertical lines indicate where zero-temperature masses are equal to respective thermal masses for the $h$ field (solid line), for the $\f$ field (long-dashed line) and for the $X$ boson (short-dashed line).\label{fig:comparison-potentials-SU2}}
\end{figure}

As a next step, we compare the improved and unimproved potentials in the SU(2)cSM model.  We now compare three different approaches to the thermal effective potential: \textit{(i)} the effective potential with thermal masses included (see eqs.~\eqref{eq:one-loop}, \eqref{eq:V-eff}, \eqref{eq:thermal-pot} and \eqref{eq:thermal-mass-first}--\eqref{eq:thermal-mass-last}) and using the high-temperature expansion for the thermal functions (see appendix~\ref{app:high-T}); \textit{(ii)} the same potential as in the previous point but with full thermal functions evaluated numerically with the use of~\cite{Fowlie:2018}; \textit{(iii)} the potential improved with the use of the gap equation. As a representative example we use the same benchmark point as before (number 7 from table~\ref{tab:benchmark}). The results for different temperatures are presented in figure~\ref{fig:comparison-potentials-SU2}. For the sake of legibility we plot the potentials in one dimension, along $\f$ with $h=0$. As we have seen in the previous section, from the point of view of the phase transition the direction along the $\f$ field is more important. We again include lines showing where the zero-temperature masses are equal to the thermal masses, for the $h$ field (solid line), for the $\f$ field (long-dashed line) and for the $X$ boson (short-dashed line). In the lower panel the solid line is out of the plot range.

Let us first focus on the effective potential using high-temperature expansion of the thermal functions. While staying close to the remaining two around the local maximum at higher temperatures, it clearly diverges around the minimum, regardless of the temperature. This clearly shows the necessity of implementing full thermal functions in practical computations.

Let us now turn to the comparison of the cases \textit{(ii)} and \textit{(iii)}, i.e.\ the daisy resummation and the gap-equation resummation. One can note that there is virtually no difference between the two potentials. If one zooms in, some differences can be seen but only at sub percent level (the difference that one can note in the lower right panel is still at sub percent level, it may seem larger than in the other plots because the scale on the vertical axis is different), which holds also for nonzero values of $h$. This can be surprising, given the results for the scalar theory we analysed before, where we observed a difference between the improved potential and the potential with thermal masses. This can be understood as one realises that in the case of SU(2)cSM we treat the contributions from the gauge bosons in the same way in both approaches of \textit{(ii)} and \textit{(iii)}, namely we use standard thermal masses for them. Remembering that the gauge contributions can be very important, especially in conformal models (see e.g.\,ref.~\cite{Chataignier:2018RSB}) one realises that the effect of improvement in the scalar sector is obscured by the contributions from gauge bosons. 

\begin{figure}[t]
\center
\includegraphics[height=.35\textwidth]{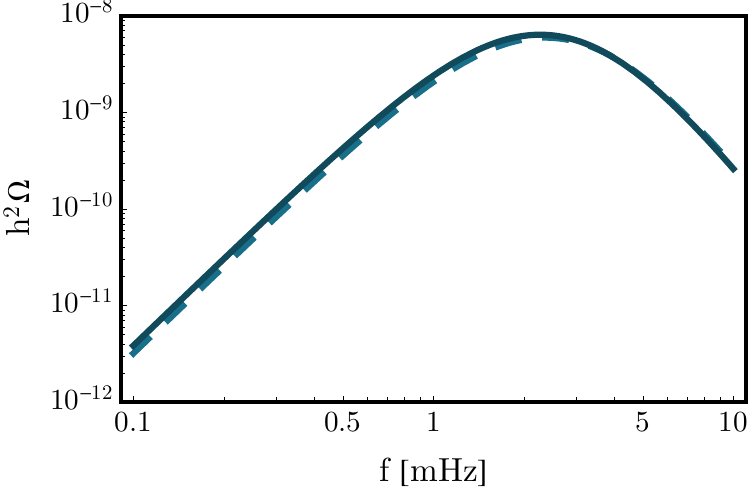}
\caption{Comparison of the GW signals following from the gap-improved potential (solid line) and the potential with daisy resummation (dashed line). \label{fig:GW-improved}}
\end{figure}

This suggests that our current method is not capable of significantly improving the results obtained from the effective potential, such as tunnelling rates or GW spectra. To verify this statement we computed the expected GW signal using the improved potential, following the same procedure as before. The result for our benchmark point, together with the result obtained from the unimproved potential can be seen in figure~\ref{fig:GW-improved}. As expected the difference is negligible, well within the numerical uncertainty.

The results presented in this section show that no gain in accuracy can be expected from improved resummation of scalar contributions to the thermal effective potential. Nonetheless, they show the potential of such resummations and indicate that further developments in the gauge sector should be undertaken.

\section{Conclusions\label{sec:conclusions}}

After the first observations of gravitational waves~\cite{Abbott:2016, Abbott:2016-2, Abbott:2017, Abbott:2017-2} a new channel of observation of the Universe was established. With the forthcoming gravitational wave detectors it will be possible to probe physics beyond the limits known before.  A prime example is the phase transition in the early Universe, which has been thoroughly studied in the context of baryogenesis and  but has not been experimentally tested.

Of course, to make use of the possibilities offered by the gravitational wave detectors, the signal predicted by theoretical models must be strong enough. In this paper we show that a~phase transition predicted by the classically conformally symmetric model with a~new SU(2) gauge group and a~scalar doublet   proceeds after large super-cooling, i.e.\ the temperature at which the transition occurs (the nucleation temperature) is much lower than the critical temperature, at which it becomes possible. This results in a large latent heat release and  a~very strong imprint in the gravitational field, well within the sensitivity range of LISA~\cite{LISA, Caprini:2015}. This can make this model, and possibly other models with classical conformal symmetry, a~great benchmark model for this detector.

Apart from the interesting gravitational-wave phenomenology, the SU(2)cSM possesses other interesting features --- it has a classical conformal symmetry which alleviates the hierarchy problem, all the masses are produced radiatively, it is perturbative (no Landau poles) and stable up to the Planck scale and it possesses a DM candidate. Last but not least, the model has few free parameters --- if apart from fixing the Higgs mass and VEV one also requires the correct DM relic abundance, there is just one free parameter, which makes the model highly predictive.  All these advantages call for a thorough and accurate treatment.

Unfortunately, the computation of the GW spectra resulting from strong first order phase transitions is plagued with many uncertainties. These include our ignorance about the dynamics of bubble formation and heat transfer in the plasma, the limited range of numerical simulations used to derive the fitting formulas, the difficulties with finding the bubble profiles in the full two-dimensional space, as well as complications with the rigorous treatment of the finite-temperature effective potential (or, more general, effective action). In the present paper we focused on improving  the last two issues.

We propose a way of extending the validity of the thermal effective potential. We introduce a resummation based on the gap equation for the improved propagator, inspired by the 2PI formalism. It allows to resum corrections to the propagator beyond the scope of the daisy resummation and does not make use of the high-temperature expansion. As a~starting point, we limit the resummation to the scalar sector of the theory, to keep the numerical procedure manageable. We have checked, however, that applying this resummation to a pure scalar theory or to SU(2)cSM does not yield significant modification of the potential or the results for the gravitational wave spectra. This implies that, either the daisy resummation works surprisingly well, or, more likely, that further developments in the resummation techniques are needed, in particular in the gauge sector.  Further on, one should use full 2PI effective action and also consider including renormalisation group improvement.

Another important observation of the present work, which can be well valid beyond the context of the SU(2)cSM considered here, is about the dynamics of the phase transition. We have shown that, at least for the benchmark points considered in this paper, the tunnelling between the vacua proceeds along the direction of the new scalar field $\f$ and not along the direction linking the global and the local minimum, which naively is the ``shortest path''. This implies that the commonly employed method for studying radiative symmetry breaking, the Gildener-Weinberg method, which reduces the potential to a one-dimensional function along the direction between the minima, cannot reproduce the correct picture of symmetry breaking. We have checked that the GW signal derived with the use of the Gildener-Weinberg method gives reasonable results, however they are different  than the ones obtained from the full two-dimensional potential. Moreover, we have checked that considering the potential solely along the direction of the new scalar field seems to be a reliable approximation.

To sum up, the SU(2)cSM model analysed in the present paper provides an exciting possibility of being tested with the future gravitational wave detectors. In order to benefit from this possibility the most we should make the effort of providing reliable and possibly precise theoretical predictions. This paper aims at making a step in this direction.

\section*{Acknowledgments}

B.\,{\'S.} is grateful to J.\,M.\,No for the enlightening discussion on gravitational waves during the Workshop on Multi-Higgs Models in Lisbon and to I.\,P.\,Ivanov and H.\,Haber for a discussion and suggestions regarding the tunnelling during the Harmonia meeting in Warsaw. We thank K.\,Olum and A.\,Masoumi for their explanations regarding AnyBubble and  A.\,Fowlie for his help with the numerical implementation of the thermal functions. T.\,P. and B.\,{\'S}. acknowledge funding from the D-ITP consortium, a~program of the NWO that is funded by the Dutch Ministry of Education, Culture and Science (OCW). This work is part of the research programme of the Foundation for Fundamental Research on Matter (FOM), which is part of the Netherlands Organisation for Scientific Research (NWO). B.{\'S}. acknowledges the support from the National Science Centre, Poland, through the HARMONIA project under contract UMO-2015/18/M/ST2/00518 (2016-2019).

\appendix
\section{Details of the  finite-temperature computations}

\subsection{Thermal masses for gauge bosons and scalars\label{app:thermal-masses}}
The thermal corrections to the mass matrix of the two scalars in eq. \eqref{eq:tree-level-mass-matrix} are given at leading order in the high-temperature expansion by
\begin{align*}
\begin{pmatrix}
m_{hh,T}^2 & m_{h \varphi, T}^2 \\ m_{\varphi h,T}^2 & m_{\varphi \varphi,T}^2
\end{pmatrix} = &\begin{pmatrix}
m_{hh}^2 & m_{h \varphi}^2 \\ m_{\varphi h}^2 & m_{\varphi \varphi}^2
\end{pmatrix} \notag \\ &+
\begin{pmatrix}
\qty(\frac{1}{4}\lambda_1 + \frac{1}{24} \lambda_2 + \frac{3}{16} g^2 + \frac{1}{16} g'^2+\frac{1}{4} y_t^2)T^2 & 0 \\
0 & \qty(\frac{1}{4}\lambda_3 + \frac{1}{24} \lambda_2 + \frac{3}{16} g_x^2 )T^2
\end{pmatrix}.
\end{align*}
The thermal masses of the two scalars are obtained by diagonalising this matrix. The factors of the scalar couplings differ from the results in the literature (e.g. ref. \cite{Carrington:1992}) since we did not include the effect of the Goldstone bosons in our computations.

At leading order only the longitudinal gauge bosons acquire a thermal mass. In the gauge field basis ($A_\mu^a$ and $B_\mu$) before electroweak symmetry breaking the longitudinal thermal masses read~\cite{Carrington:1992, Sannino:2015}
\begin{align*}
M_L^2(h,T) = \frac{h^2}{4} \begin{pmatrix}
g^2 && 0 && 0 && 0 \\ 0 && g^2 && 0 && 0 \\ 0 && 0 && g^2 && - g g' \\  0 && 0 && -g g' && g'^2
\end{pmatrix} +\frac{11}{6} T^2 \begin{pmatrix}
g^2 && 0 && 0 && 0 \\ 0 && g^2 && 0 && 0 \\ 0 && 0 && g^2 && 0 \\  0 && 0 && 0 && g'^2
\end{pmatrix}.
\end{align*}
Transforming the above matrix to the $W_\mu^+$, $W_\mu^-$, $Z_\mu$ and $A_\mu$ basis, we obtain the thermal masses of the longitudinal $W$, $Z$ and photon \cite{Rose:2015}
\begin{align}
M^2_{W_L} &= m_W^2(h) + \frac{11}{6} g^2 T^2, \label{eq:thermal-mass-first}\\
M^2_{Z_L} &= \frac{1}{2} m_Z^2(h) +\frac{11}{12} \frac{g^2}{\cos^2(\theta_W)} T^2 + \frac{\Delta}{2},\\
M^2_{\gamma_L} &=\frac{1}{2} m_Z^2(h)+  \frac{11}{12} \frac{g^2}{\cos^2(\theta_W)} T^2 - \frac{\Delta}{2}, \\
\Delta^2 &= m_Z^4(h) + \frac{11}{3} \frac{g^2 \cos^2(2\theta_W)}{\cos^2(\theta_W)}\qty[m_Z^2(h) + \frac{11}{12} \frac{g^2}{\cos^2(\theta_W)} T^2] T^2.\label{eq:thermal-mass-last}
\end{align}
\par
For the gauge bosons of the hidden gauge group, the thermal masses resemble the ones for the $W$ bosons but with the gauge coupling $g_X$. However, the factor $11/6$ is replaced by $5/6$ since the $X$ gauge bosons do not couple to any fermions. The coupling of the SU(2)$_X$ gauge group is relatively large, therefore, the high-temperature result may not be sufficiently suppressed by the coupling constant at lower temperatures/large field values. To account for  a smooth transition between high and low temperature (in the absence of a reliable analytical solution), we use for the longitudinal mass of the $X$ gauge boson the expression
\begin{align*}
M_{X_L}^2 =  m_X^2(\varphi) + \frac{1}{2} \qty(\tanh(a\qty(-\varphi +  \frac{2 T}{g_X}))+1)  \frac{5}{6} g_X^2 T^2 .
\end{align*}
The parameter $a$ regulates the smoothness of the transition of both limits and is set to $a=0.01\g^{-1}$ in our computations. This is a crude approximation to the physical mass of the $X$ gauge boson. To account for a more accurate thermal mass for the $X$ gauge boson, it is desirable to generalise the gap equation approach to gauge bosons which is left for future research.

 Corrections to the top quark mass are not necessary since fermions do not suffer from IR divergences due to their finite zero Matsubara mode~\cite{Espinosa:1992-2}.

\subsection{High-temperature expansion of the thermal functions\label{app:high-T}}
 The integrals in the definition of the $J_B$ and $J_F$ functions, eq.~\eqref{eq:thermal-functions}, can be solved analytically if a high temperature expansion is performed, yielding
\begin{align}
J_B(y^2) = & - \frac{\pi^4}{45} + \frac{\pi^2}{12} y^2 - \frac{\pi}{6} \left( y^2 \right)^{3/2} - \frac{1}{32}y^4 \log \frac{y^2}{a_B} \notag \\*
&- 2 \pi^{7/8} \sum_{m=1}^\infty \frac{(-1)^m}{(m+2)!} \Gamma\qty(m+\frac{1}{2}) \left(\frac{y^2}{4 \pi^2}\right)^{m+2} \zeta(2m+1),\notag \\
 J_F(y^2) = & \frac{7 \pi^4}{360}-\frac{\pi^2}{24}y^2 - \frac{1}{32} y^4 \log \frac{y^2}{a_f} \notag \\*
 & - \frac{\pi^{7/2}}{4} \sum_{l=1}^\infty (-1)^l \frac{\zeta(2l+1)}{(l+1)!}(1-2^{-2l-1})\Gamma \qty(l+\frac{1}{2}) \qty(\frac{y^2}{\pi^2})^{l+2}, \notag
\end{align}
which is valid for $y^2 = (m/T)^2 \ll 1$ and
\begin{align}
a_B= &16 \pi^2 \exp(\frac{3}{2}-2 \gamma_E), \notag\\
 a_f = & \pi^2 \exp(\frac{3}{2}-2\gamma_E).\notag
\end{align}

\subsection{Discussion of the values of the $\beta$ parameter\label{app:beta}}

\begin{figure}[t]
\center
\includegraphics[height=.3\textwidth]{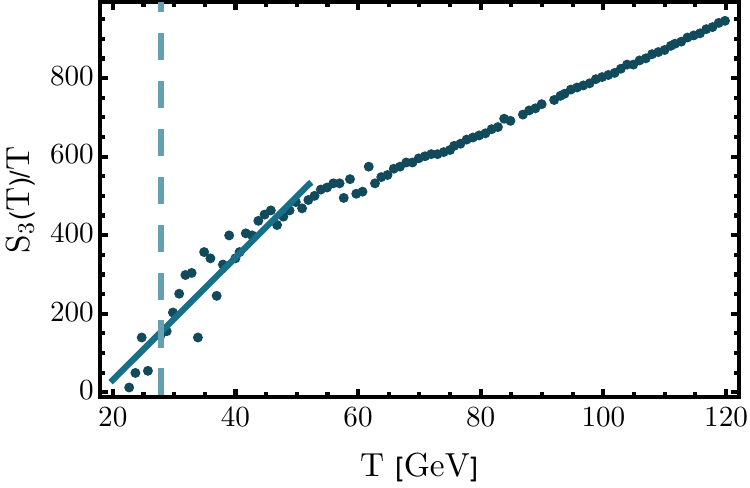}
\includegraphics[height=.3\textwidth]{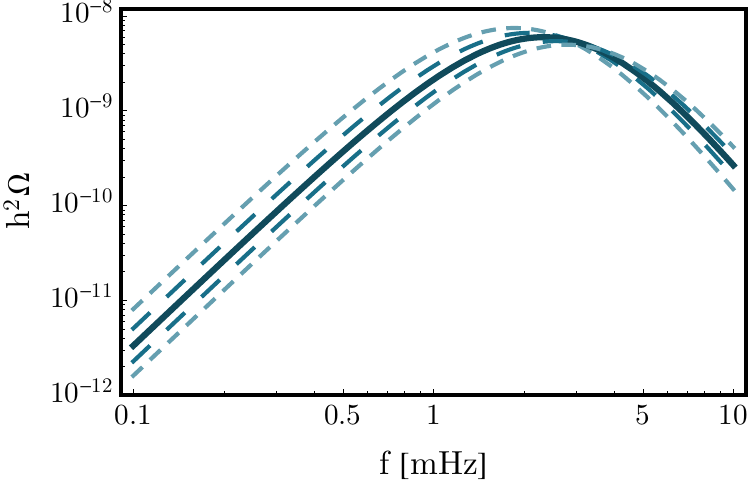}
\caption{Left panel: the values of $S_3(T)/T$ as a function of $T$, together with the line fitted around the nucleation temperature. The dashed line indicates the nucleation temperature. Right panel: The GW signal for the benchmark point with the central value of $\beta/H_*$ (solid line), $\beta/H_*$ modified by $\pm10\%$ (long-dashed line) and by $\pm20\%$ (short-dashed line). \label{fig:beta}}
\end{figure}

The inverse time scale of the transition, $\beta/H_*$ is defined as in eq.~\eqref{eq:beta}. In the numerical procedure that we use, we find the values of the Euclidean action, $S_3$ for different temperatures numerically, using the code AnyBubble. Therefore, we have to find the derivative of eq.~\eqref{eq:beta} numerically. We do this by fitting a straight line to the points around the nucleation temperature. A plot of the values of $S_3(T)/T$ as function of the temperature is shown in figure~\ref{fig:beta} (left panel) together with the fitted line. The slope of the line is taken as the value of the derivative. It can be seen, that the spread of the points, originating from numerical inaccuracies in evaluation of $S_3$, around the nucleation temperature (marked by the dashed line) is significant, which can be caused, at least to some extend, by the shallowness of the minimum. Moreover, for some temperatures the code does not complete the computation. Therefore, we cannot take too few points to fit the line, which in turn introduces additional uncertainty  --- the value depends on the arbitrary choice of the number of points to include. We estimate the error in determining $\beta/H_*$ to be at the level of up to around $20\%$. In the one-dimensional cases that we consider (the Gildener-Weinberg method and the tunnelling along the $h=0$ direction) we obtain fewer points around the nucleation temperature, therefore, it is more difficult to estimate the error.

To verify how large impact on the predictions of the GW signal this error can have in figure~\ref{fig:beta} (right panel) we show the GW signal for the benchmark point analysed in the main text (number 7 from table~\ref{tab:benchmark}) with $\beta/H_*$ having the value given in table~\ref{tab:benchmark}, the value modified by $\pm10\%$ (long-dashed line) and $\pm 20\%$ (short-dashed line). It is clear that for precise predictions to be confronted with the experimental results, one needs to reduce this uncertainty. However, the error does not invalidate the main conclusion --- that the GW signal is strong, with the peak frequency $\mathcal{O}$(1mHz).

A possible way to improve the accuracy of $\beta/H_*$ is to use the result of the present analysis, that the tunnelling only proceeds along the $\f$ direction, and solve the bounce equation using simple undershooting--overshooting method in one dimension to find the Euclidean action with better accuracy. Another possibility is to work on the numerical implementation of the thermal effective potential that would be more compatible with the AnyBubble code.
 
\section{Gildener-Weinberg method\label{app:Gildener-Weinberg}}
A method of dealing with multi-variable potentials was proposed by Gildener and Weinberg in ref.~\cite{Gildener:1976}. It is based on an assumption that generally the tree-level potential dominates over the one-loop correction. However, there exists a renormalisation scale at which the tree-level potential develops a flat direction, along which its value is zero and along this direction the one-loop correction becomes the leading contribution  and should be analysed. This reduces the dimensionality of the problem to one,  greatly simplifying the problem.  A discussion of applicability of this method at zero temperature can be found in ref.~\cite{Chataignier:2018RSB}. We employ this approach in the present work in order to compare the results to the ones obtained from the full one-loop potential and argue that the GW approach is not a reliable tool for studying phase transitions.

For the scalar potential of the SU(2)cSM, eq.~\eqref{eq:Vtree}, a flat direction is attained at the GW scale $\mgw$ defined through
\be
4 \la\lc - \lb^2\big|_{\mgw} = 0.\notag
\ee
Assuming that the direction from the origin to the minimum is not significantly changed by including the loop corrections one obtains through minimisation of the tree-level potential at the GW scale the following relation between the VEVs of the scalar fields
\be
 w^2=-\frac{\lb}{2\lc}v^2=-\frac{2\la}{\lb}v^2,\notag
 \ee
where $w=\langle \f\rangle$, $v=\langle h\rangle$.
We can define the angle to the minimum as
\be
\tan \alpha =\frac{w}{v}.\notag
\ee
The tree-level mass matrix (the Hessian of the tree-level potential) has one vanishing eigenvalue, corresponding to the flat direction, and the other one given by
\be
M_1^2=(2\la-\lb)v^2=-\lb\rho^2, \label{eq:GW-tree-mass}
\ee
where $\rho^2=v^2+w^2$. Due to the presence of the flat direction, the angle $\alpha$ defined above diagonalises the tree-level mass matrix. 

Inclusion of the loop correction lifts the flat direction and generates mass for the particle that is massless at tree level. The mass reads
\be
 M_2^2=\frac{8\mathbb{B}}{\rho^2},\quad\quad \mathbb{B}(\mu, \lambda_j,\phi)=\frac{1}{64 \pi^2}\sum_{a}n_a M_a^4(\phi),\notag
 \ee
 where $\phi$ denotes the field variable along the tree-level flat direction and the masses are computed at the GW scale, thus the Goldstone bosons do not contribute (they are massless along the tree-level minimum) and the remaining scalar contributes with mass given in eq.~\eqref{eq:GW-tree-mass}, with $\rho$ replaced by $\phi$.

\bibliography{conformal-bib+GW}

\end{document}